\documentclass[5p]{elsarticle}
\usepackage[english]{babel}
\usepackage[utf8]{inputenc}	%Hvis du benytter windows i stedet for linux, så skift utf8 ud med latin1. Tillader danske tegn.
\usepackage[T1]{fontenc}	%Tillader danske tegn
\usepackage{graphicx}
\usepackage{siunitx}	%Bruges \SI{<tal>}{<enhed>}, \si{<enhed>} eller \num{<tal>}.
\usepackage{url}		 %bruges til at formattere url'er... kan sagtens udelades.
\addto\captionsenglish{\renewcommand{\figurename}{Fig.}}

\newcommand{\fref}[1]{\figurename~\ref{#1}}

\usepackage{hyperref}
\hypersetup
{   pdfsubject={Paper},
	pdfauthor={Jonas Refsgaard}
    pdftitle={Measurement of the branching to the 9.6MeV state in 16O in the beta-decay of 16N},
    pdfstartview=FitH,
    colorlinks=true}
    
\makeatletter
\begingroup
\catcode`\_=\active
\protected\gdef_{\@ifnextchar|\subtextup\sb}
\endgroup
\def\subtextup|#1|{\sb{\textup{#1}}}
\AtBeginDocument{\catcode`\_=12 \mathcode`\_=32768 }
\makeatother

\usepackage[]{csquotes} %Danske citationstegn. \enquote{}
\usepackage{bm}
\usepackage{amsmath}

\newcommand{\nitrogen}{$^{16}\mathrm{N}$}

\begin{document}

%\title{Branching ratio for the \texorpdfstring{$\boldsymbol\beta$}{beta}-delayed \texorpdfstring{$\boldsymbol\alpha$}{alpha}-decay of %\texorpdfstring{$^{16}\mathbf{N}$}{16N}}
\title{Measurement of the branching ratio for \texorpdfstring{$\beta$}{beta}-delayed \texorpdfstring{$\alpha$}{alpha} decay of \texorpdfstring{$^{16}\mathrm{N}$}{16N}}
%\title{Measurement of the $\beta\alpha$ branching ratio of \texorpdfstring{$^{16}\mathrm{N}$}{16N}} %%%OLI

\author[au]{J.~Refsgaard\corref{cor}}
\ead{jr@phys.au.dk}
\author[au]{O.\,S.~Kirsebom}
\author[kvi]{E.\,A.~Dijck}
\author[au]{H.\,O.\,U.~Fynbo}
\author[au]{M.\,V.~Lund}
%\author[kvi]{M.\,Nu\~{n}es.~Portela}
\author[kvi]{M.\,N.~Portela} %%%JR
\author[kul]{R.~Raabe}
\author[kul]{G.~Randisi}
\author[kul]{F.~Renzi}
\author[kul]{S.~Sambi}
\author[kvi]{A.~Sytema}
\author[kvi]{L.~Willmann}
\author[kvi]{H.\,W.~Wilschut}
\cortext[cor]{Corresponding author}
\address[au]{Department of Physics and Astronomy, Aarhus University, DK-8000 Aarhus, Denmark}
\address[kul]{KU Leuven, Instituut voor Kern- en Stralingsfysica, B-3001 Leuven, Belgium}
\address[kvi]{Van Swinderen Institute for Particle Physics and Gravity, University of Groningen, NL-9747 AG Groningen, The Netherlands}


\begin{abstract}
While the $^{12}\mathrm{C}(\alpha,\gamma)^{16}\mathrm{O}$ reaction plays a central role in nuclear astrophysics, the cross section at energies relevant to hydrostatic helium burning is too small to 
be directly measured in the laboratory. The $\beta$-delayed $\alpha$ spectrum of $^{16}$N can be used to constrain the extrapolation of the E1 component of the $S$-factor; however, with this approach the resulting $S$-factor becomes strongly correlated with the assumed $\beta\alpha$ branching ratio. We have remeasured the $\beta\alpha$ branching ratio by implanting $^{16}$N ions in a segmented Si detector and counting the number of $\beta\alpha$ decays relative to the number of implantations. Our result, \num{1.49 \pm 0.05 e-5}, represents a \SI{24}{\percent} increase compared to the accepted value and implies an increase of $\SI{\approx 13}{\percent}$ in the extrapolated $S$-factor.

\end{abstract}

\maketitle

\section{Introduction}
\subsection{The astrophysical \texorpdfstring{$S$}{S}-factor of \texorpdfstring{$^{12}\mathrm{C}(\alpha$,$\gamma)^{16}\mathrm{O}$}{12C(a,g)16O}.}\label{sec:s-factor}%
The rate of the $^{12}\mathrm{C}(\alpha$,$\gamma)^{16}\mathrm{O}$ reaction, relative to that of the triple-$\alpha$ reaction, regulates the relative production of carbon (C) and oxygen (O) during hydrostatic helium burning in stars, and by doing so has great significance for the field of nuclear astrophysics. Not only is the C/O ratio at the end of helium burning directly reflected in the observed elemental abundances, it also has profound influence on subsequent stellar evolution: It affects the nucleosynthesis of medium-mass and $s$-process-only nuclei~\cite{west2013}, long-lived $\gamma$-ray emitters~\cite{tur2010}, and $\nu$-process nuclei~\cite{austin2014}, it affects the chemical composition of white-dwarf stars and thus explosion models of type-IA supernovae~\cite{thielemann2004}, and it affects the mass of the remnant left behind by core-collapse supernovae~\cite{woosley2003}. %

At the centre-of-mass energies relevant to hydrostatic helium burning, $E_{\textrm{c.m.}} \approx \SI{0.3}{\mega\electronvolt}$, the cross section of the $^{12}$C($\alpha$,$\gamma$)$^{16}$O reaction is too small to be measured directly in the laboratory. Indeed, the lowest point measured so far is $E_{\textrm{c.m.}}=\SI{0.89}{\mega\electronvolt}$~\cite{hammer2005}. 
The prospects of extending these measurements down to \SI{0.3}{\mega\electronvolt} within the foreseeable future are incredibly slim, as it would require an improvement of more than 5 orders of magnitude in experimental sensitivity. 

A precise and accurate extrapolation to the astrophysically relevant energies can be made only by including complementary data obtained by indirect methods. %
At \SI{0.3}{\mega\electronvolt} the $S$-factor of the $^{12}$C($\alpha$,$\gamma$)$^{16}$O reaction is dominated by E1 and E2 resonant capture to the ground state via the tails of the subthreshold $1^-$ and $2^+$ states at 7.12 and \SI{6.92}{\mega\electronvolt}, with smaller contributions from cascade transitions~\cite{buchmann2006}. %
$\gamma$-ray angular distributions measured at higher energies are used to determine the relative contribution of the E1 and E2 components. %

The extrapolation of the E2 component may be constrained using $d$-wave phase-shift data from ($\alpha$,$\alpha$) elastic scattering experiments. Using this approach, Tischhauser {\it et al.}\ have obtained $S_{\textrm{E}2}(0.3)=53^{+13}_{-18}\;\si{\kilo\electronvolt\barn}$~\cite{tischhauser2002}, which is also the value quoted by Buchmann and Barnes in their 2006 review article~\cite{buchmann2006}. Taking advantage of new $(\alpha,\gamma)$ capture data, Sayre {\it et al.}\ have recently obtained the more precise value of $S_{\textrm{E}2}(0.3)=62^{+9}_{-6}\;\si{\kilo\electronvolt\barn}$~\cite{sayre2012}. 
Complementary constraints on $S_{\textrm{E}2}(0.3)$ have been obtained from sub-Coulomb $\alpha$-transfer reactions~\cite{brune1999} and a $\gamma$-cascade experiment~\cite{matei2008}. An alternative approach based on Coulomb dissociation has also been explored~\cite{fleurot2005}. % F. Fleurot et al., Phys Lett. B 615 (2005) 167--174.

The extrapolation of the E1 component is only weakly constrained by the $p$-wave phase-shift data~\cite{tischhauser2002, buchmann2006}. The precision and accuracy of the extrapolation can be significantly improved by including data from the $\beta\alpha$ decay of $^{16}$N~\cite{barker1971}. %
Using this approach, Azuma {\it et al.}\ have obtained $S_|E1|(0.3)=\SI{80 \pm 20}{\kilo\electronvolt\barn}$~\cite{azuma1994}, which is also the value quoted by Buchmann and Barnes~\cite{buchmann2006}. More recently, Tang {\it et al.}\ have obtained $S_|E1|(0.3)=\SI{84 \pm 21}{\kilo\electronvolt\barn}$~\cite{tang2010}. %
Other authors have used data from sub-Coulomb $\alpha$-transfer reactions to obtain complementary constraints on $S_{\textrm{E}1}(0.3)$~\cite{brune1999}.

Cascade transitions via the four bound excited states in $^{16}$O were previously believed to contribute anywhere between 3 and \SI{32}{\kilo\electronvolt\barn} to the total $S(0.3)$-factor~\cite{buchmann2006}, but new data~\cite{sayre2012, avila2015} appear to have constrained the their contribution to \SI{11 \pm 3}{\kilo\electronvolt\barn}.
%
% from an $\alpha$-transfer reaction experiment at sub-Coulomb energies~\cite{avila2015} has determined the contribution of the $0^+$, $3^-$ cascade transitions to be between $2.1\pm0.3$ and $5.8\pm0.5$~keV~b (=4.0^{+2.3}_{-2.2}), and new data from a $\gamma$-$\gamma$ coincidence measurement~\cite{matei2008} has determined the contribution of the $2^+$ cascade transition to be $7.1\pm 1.6$~keV~b. Since the contribution of the $1^-$ cascade transition is believed to be very small, 0.3~keV~b, the summed contribution of the cascade transitions is 
%
Combining this with $S_|E2|(0.3)=62^{+9}_{-6}\;\si{\kilo\electronvolt\barn}$~\cite{sayre2012} and $S_|E1|(0.3)=\SI{84 \pm 21}{\kilo\electronvolt\barn}$~\cite{tang2010}, one obtains a total $S$-factor of $S(0.3)=\SI{157 \pm 23}{\kilo\electronvolt\barn}$. %
Taken at face value, the new results of Refs.~\cite{sayre2012, avila2015} thus imply a significant change in the error budget, with the E1 ground-state capture now making by far the largest contribution (\SI{13}{\percent}) to the overall error, while the E2 ground-state capture and the cascade transitions only make a modest contribution (\SI{6}{\percent}). %
A precision of \SI{10}{\percent} has long been desired by astrophysical modellers~\cite{weaver1993, woosley2002, woosley2007}. % which would allow them to constrain stellar-model uncertainties such as convection.
This provides strong motivation for reducing the uncertainty on $S_|E1|(0.3)$.

It should be noted that the uncertainty on the extrapolated $S$-factor is a subject of strong debate. Several authors have argued that the experimental data is compatible with two different values of $S_|E1|(0.3)$, a high value around \SI{80}{\kilo\electronvolt\barn} and a low value around \SI{10}{\kilo\electronvolt\barn}, see, {\it e.g.}, Refs.~\cite{hale1997, gialanella2001}. Similarly, it has been argued~\cite{gai2013} that two solutions exist for $S_{\textrm{E}2}(0.3)$, a high value around \SI{150}{\kilo\electronvolt\barn} and a low value around \SI{60}{\kilo\electronvolt\barn}. %
Furthermore, Tang~{\it et al.} have shown~\cite{tang2010} that the $S_|E1|(0.3)$ value obtained from the simultaneous analysis of phase-shift, capture and $\beta\alpha$-decay data is reduced by 20--\SI{30}{\percent} if the old phase-shift data of Ref.~\cite{plaga1987} are replaced with the more recent phase-shift data of Ref.~\cite{tischhauser2002, tischhauser2009}.
These observations stand in stark contrast to the very precise value of $S(0.3)=161\pm 19$(stat)$^{+8}_{-2}$(sys)\si{\kilo\electronvolt\barn} recently reported by Sch\"urmann {\it et al.}~\cite{schurmann2012} based on a global analysis of a selected sample of ``world data''. The data selection criteria adopted by Sch\"urmann {\it et al.}\ have since been criticised by several authors~\cite{gai2013, brune2013}. %

\subsection{The \texorpdfstring{$\beta$}{beta}-delayed \texorpdfstring{$\alpha$}{alpha} decay of \texorpdfstring{$^{16}\mathrm{N}$}{16N}}\label{sec:br}%
The focus of the present work is on reducing the uncertainty on $S_{\textrm{E}1}(0.3)$ by improving the experimental determination of the $\beta\alpha$ decay of $^{16}$N. 
The only $\alpha$-decaying state in $^{16}$O that is appreciably fed in the $\beta$ decay of $^{16}$N is the $1^-$ state at \SI{9.6}{\mega\electronvolt}, and as a result this state dominates the $\alpha$ spectrum. So much, in fact, that earlier determinations of its $\beta$ branching ratio, $b_{\beta}(9.6)$, have been obtained by measuring the total $\beta\alpha$ branching ratio, $b_{\beta\alpha}$, and assuming $b_{\beta}(9.6) = b_{\beta\alpha}$, which is also the approach that we shall follow.
There is, however, also a small contribution from the high-energy tail of the subthreshold $1^-$ state at \SI{7.12}{\mega\electronvolt}, and hence a careful measurement of the $\alpha$ spectrum can be used to constrain the $\alpha$ width of the \SI{7.12}{\mega\electronvolt} state, which dominates the E1 component of the ground-state capture at the astrophysically relevant energies. %
%
%The influence of the \SI{7.12}{\mega\electronvolt} state on the $\alpha$ spectrum is amplified by the smallness of the $\beta$ branching ratio of the \SI{9.6}{\mega\electronvolt} state, $b_{\beta}(9.6) \sim 1\times 10^{-5}$, relative to the $\beta$ branching ratio of the \SI{7.12}{\mega\electronvolt} state, $b_{\beta}(7.12) \sim 5\times 10^{-2}$. % 
%
The influence of the \SI{7.12}{\mega\electronvolt} state on the $\alpha$ spectrum is enhanced by its large $\beta$ branching ratio, $b_{\beta}(7.12) \sim 5\times 10^{-2}$, relative to the $\beta$ branching ratio of the \SI{9.6}{\mega\electronvolt} state, $b_{\beta}(9.6) \sim 1\times 10^{-5}$. %%%JR: Paragraph could be omitted altogether if pressed for space.

While consensus appears to have been established concerning the shape of the experimental $\alpha$ spectrum~\cite{tang2010, buchmann2009}, an improved measurement of $b_{\beta}(9.6)$ is needed to fix the absolute normalisation of the spectrum~\cite{buchmann2009}. % 
%

%%% OK - smaa aendringer i afsnittet nedenfor %%%% 
Azuma {\it et al.} have shown that the $\alpha$ width of the \SI{7.12}{\mega\electronvolt} state, and hence the value of $S_{\textrm{E}1}(0.3)$, obtained from an $R$-matrix analysis of the $\beta$-delayed $\alpha$ spectrum (simultaneously constrained by elastic-scattering and capture data) is strongly correlated with the relative feeding of the two $1^-$ states. In particular, a $\pm \SI{9}{\percent}$ uncertainty in the ratio $b_{\beta}(9.6)/b_{\beta}(7.12)$ results in a $\pm \SI{6}{\kilo\electronvolt\barn}$ uncertainty on $S_|E1|(0.3)$~\cite{azuma1994}. 

We present the following, qualitative analysis of how a change in the assumed $b_{\beta}(9.6)$ affects the calculated $S_{\textrm{E}1}(0.3)$: Consider a level $\lambda$ with a $\beta$ feeding amplitude, $B_{\lambda}$, and a reduced $\alpha$ width, $\gamma_{\lambda\alpha}$. The contribution of that level to the total $\alpha$ spectrum is proportional to $(B_{\lambda}\gamma_{\lambda\alpha})^2$ (following the notation of Barker and Warburton~\cite{barker1988}). In order to scale the spectrum by some factor, $f$, we must, if the general shape of the spectrum is to be preserved, scale $(B_{\lambda}\gamma_{\lambda\alpha})^2$ for all contributing levels by the same factor $f$. Since $B_{\lambda}$ for the \SI{7.12}{\mega\electronvolt} state has been determined experimentally with high precision~\cite{tang2010}, any change in the contribution of this level to the $\alpha$ spectrum must involve a change in its $\gamma_{\lambda\alpha}$. From these considerations we conclude that a change in the absolute scale of the $\alpha$ spectrum, \textit{i.e.} $b_{\beta}(\num{9.6}) \rightarrow fb_{\beta}(\num{9.6})$, must be accompanied by a change $\gamma_{\lambda\alpha}^2 \rightarrow f \gamma_{\lambda\alpha}^2$ for the \SI{7.12}{\mega\electronvolt} state, which, to a reasonable approximation, leads to $S_|E1|(0.3) \rightarrow f S_|E1|(0.3)$, consistent with the numerical result of Azuma {\it et al.}

We proceed by reviewing what is known experimentally about $b_{\beta}(7.12)$ and $b_{\beta}(9.6)$: Tang {\it et al.}\ have recently determined $b_{\beta}(7.12)=\num{0.052 \pm 0.002}$, though details have not yet been published. 
%%%%%%%%%%%%%%%%%%
The TUNL evaluation~\cite{tunl16} gives $b_{\beta}(9.6)=\num{1.20 \pm 0.05 e-5}$, implying an uncertainty of \SI{6}{\percent} in the ratio $b_{\beta}(9.6)/b_{\beta}(7.12)$ and hence an uncertainty of $\pm \SI{4}{\kilo\electronvolt\barn}$ in $S_|E1|(0.3)$. However, this value of $b_{\beta}(9.6)$ is based on a single measurement that dates back to 1961~\cite{kaufmann1961}. The measurement, performed by the Mainz group, relied on the direct counting of $\beta$ particles with a Geiger-M\"uller tube. No details are given on the error estimate, though based on the large number of detected $\alpha$ particles (\num{5.0e4}) it may be concluded that systematic errors dominate. %
Slightly different values can be found in two later publications of the Mainz group: Ref.~\cite{hattig1969} gives $b_{\beta}(9.6)=\num{1.19 \pm 0.04 e-5}$ quoting Ref.~\cite{kaufmann1961} as the source, whereas Ref.~\cite{hattig1970} gives the same value, but with an inflated error, $b_{\beta}(9.6)=\num{1.19 \pm 0.10 e-5}$, quoting Ref.~\cite{alburger1959} as the source, which would appear to be a mistake because Ref.~\cite{alburger1959} makes no mention of the \SI{9.6}{\mega\electronvolt} state, let alone its $\beta$ branching ratio. %

An independent determination of $b_{\beta}(9.6)$ may be obtained from the relative feeding of the 9.6 and \SI{8.87}{\mega\electronvolt} states, determined to be $b_{\beta}(9.6)/b_{\beta}(8.87)=\num{1.00\pm0.07 e-3}$ by the Mainz group~\cite{neubeck1974}, combined with the $\beta$ branching ratio of the \SI{8.87}{\mega\electronvolt} state, given as $b_{\beta}(8.87)=\num{1.06\pm0.07 e-2}$ in the current (1993) TUNL evaluation. One thus obtains $b_{\beta}(9.6)=\num{1.06\pm0.10 e-5}$. However, the origin of the $b_{\beta}(8.87)$ value given in the current TUNL evaluation is difficult to trace. The 1986 evaluation quotes E.~K.~Warburton, private communication, as the source, while earlier evaluations give a value of \num{1.0 \pm 0.2 e-2} quoting several experiments from the 1950s as sources. The reliability of the $b_{\beta}(8.87)$ value given in the current TUNL evaluation is thus difficult for us to assess. %

Recently, Zhao {\it et al.}~\cite{zhao1993} have obtained $b_{\beta}(9.6) = \num{1.3\pm 0.3 e-5}$ using an experimental technique similar to the one described here.

\subsection{Experimental technique}
We use an experimental technique very different from that of the Mainz group to determine $b_{\beta}(9.6)$. %
We implant a mass-separated, high-energy, $^{16}$N beam in a finely segmented Si detector and measure the two ionisation signals produced by the $^{16}$N implantation and the $\alpha+{}^{12}$C decay products that follow with a half-life of \SI{7.13 \pm 0.02}{\second}~\cite{tunl16}. This allows us to determine $b_{\beta}(9.6)$ in a very straightforward manner as the number of $\alpha$ decays divided by the number of implantations, and the difficulties associated with absolute counting of $\beta$ particles are thus avoided. %
This technique has been used in several previous experiments to determine small branching ratios in the $\beta$-delayed particle decays of $^6$He at Lovain-la-Neuve~\cite{smirnov2005,raabe2009}, $^{11}$Li at TRIUMF~\cite{raabe2008}, $^{12}$B and $^{12}$N at KVI~\cite{hyldegaard2009,hyldegaard2010}, and $^8$B also at KVI~\cite{roger2012}. %
%
%In the studies of $^{11}$Li, $^{12}$B, $^{12}$N and $^8$B the implantation rate was sufficiently low to allow correlation of implantation and decay events on an individual basis. This was not possible in the study of $^6$He and has not been possible in the present study due to the combination of a very small branching ratio for $\beta$-delayed particle decay and a rather long half-life.
%
In the studies of $^{11}$Li, $^{12}$B, $^{12}$N and $^8$B it was possible to correlate the implantation and decay events on an individual basis. This was not possible in the study of $^6$He and has not been possible in the present study due to the combination of a rather long half-life and a high implantation rate necessary to obtain satisfactory statistical precision.%%%JR.
%
%In the case of $^{12}$B and $^{12}$N, previously determined $\beta\alpha$ branching ratios were found to be in error by a factor of 2~\cite{hyldegaard2009}.

\section{Experiment}

The experiment was carried out at the former Kernfysisch Versneller Instituut (KVI) in Groningen, The Netherlands. %
A primary $^{15}\mathrm{N}$ beam was accelerated by the AGOR superconducting cyclotron to an energy of $\SI{105}{\mega\electronvolt}$ and directed onto a $\mathrm{CD}_2$ gas target with a thickness of \SI{6}{\milli\gram\per\square\centi\meter}. The secondary beam emerging from the gas target consisted, among other isotopes, of $^{16}\mathrm{N}$, % with a kinetic energy of approximately \SI{80}{\mega\electronvolt\per ion}, 
produced via $(d,p)$. The TRI$\mathrm{\mu}$P dual magnetic separator~\cite{berg2006} was tuned to select $^{16}\mathrm{N}^{7+}$ ions with an energy of \SI{80}{\mega\electronvolt}. At the final focal plane \SI{83}{\percent} of the secondary beam was identified as $^{16}\mathrm{N}$. %

\begin{figure}[htbp]
\centering
%\begin{pspicture}(-3,3)(0,10)
%\psframe(-3,-0.5)(-3.2,0.5) %dE
%\psframe(1.9,-0.5)(2.1,0.5) %dsssd
%\psframe(1.4,1.1)(2.6,2.1) %NaI1
%\psframe(1.4,-1.1)(2.6,-2.1) %NaI2
%\psframe[fillstyle=solid,fillcolor=black](-0.6,0.4)(-0.4,0.7)
%\psframe[fillstyle=solid,fillcolor=black](-0.6,-0.4)(-0.4,-0.7)
%\psframe[fillstyle=solid,fillcolor=black](1.87,-0.45)(1.95,-0.30)
%\psframe[fillstyle=solid,fillcolor=black](1.87,-0.20)(1.95,-0.05)
%\psframe[fillstyle=solid,fillcolor=black](1.87,0.05)(1.95,0.20)
%\psframe[fillstyle=solid,fillcolor=black](1.87,0.30)(1.95,0.45)
%\psframe[fillstyle=solid,fillcolor=black](2.05,-0.45)(2.13,0.45)
%%\psline[linestyle=dashed](-0.6,0.4)(-0.6,-0.4)
%%\psline[linestyle=dashed](-0.4,0.4)(-0.4,-0.4)
%\psline[linestyle=dotted]{-}(-4.0,0)(2,0)
%\psline[linestyle=dotted,arrowsize=8pt]{->}(-5.0,0)(-4.0,0)
%\rput(-4.5,-0.4){Ion-beam}
%\rput(-4.12,-0.85){from separator}
%\rput(-3.1,0.9){$\Delta E$-detector}
%\rput(1.15,-0.4){DSSSD}
%%\psline[arrowsize=8pt]{->}(0.2,-1.25)(1.6,-0.3)
%\rput(-0.5,1.1){Collimator}
%\rput(2.0,1.6){NaI}
%\rput(2.0,-1.6){NaI}
%\end{pspicture}
\includegraphics[scale=1]{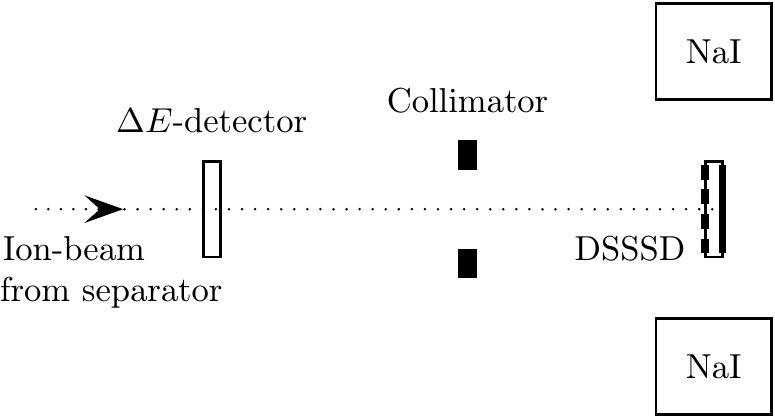}
\caption{Drawing of the detector setup (not to scale). Details on the detectors can be found in the text.}
\label{setup:fig}
\end{figure}

The detector system, sketched in \fref{setup:fig}, consisted of a \SI{60}{\micro\meter} thick circular Si detector with a diameter of \SI{18}{\milli\meter}, and a double-sided Si strip detector (DSSSD) with a thickness of \SI{78}{\micro\meter} and a surface area of $\SI{16}{\milli\meter}\times\SI{16}{\milli\meter}$. The two detectors were mounted in a telescope configuration with the circular detector serving as $\Delta E$ detector, leaving the DSSSD, thick enough to fully stop the {\nitrogen} ions, to detect their remaining energy. This type of setup provides a means to distinguish {\nitrogen} from other beam components, since the difference in stopping power can be exploited.
Lastly, two NaI scintillators were placed next to the chamber to provide $\gamma$-ray identification of {\nitrogen} during the initial beam tuning.

%The mass-separated {\nitrogen} beam has an energy of approximately \SI{80}{\mega\electronvolt} and is implanted approximately halfway into the DSSSD, thus depositing on average \SI{33}{\mega\electronvolt} in the DSSSD. %
%To avoid overheating, the DSSSD is placed on a copper mount, cooled to \SI{2}{\celsius} using a liquid cooling system. %
%

48 strips on both sides of the DSSSD, running in perpendicular directions, divide the detector into a total of 2304 pixels, each representing an active volume of approximately $\SI{300}{\micro\meter}\times\SI{300}{\micro\meter}\times\SI{78}{\micro\meter}$~\cite{sellin1992}. 
The smallness of the detection volume implies that the detector is inherently $\beta$ suppressed, {\it i.e.}, $\beta$ particles from the decay of {\nitrogen} deposit only little energy in one pixel. This has two advantages: Firstly, the distortion of the $\alpha$ spectrum due to $\beta$ summing is minimized, and secondly, the $\beta$-singles spectrum does not extend into the energy region relevant for the identification of the $\alpha$-decay branch. 

An $\alpha$ source, consisting of $^{239}\mathrm{Pu}$, $^{241}\mathrm{Am}$ and $^{244}\mathrm{Cm}$, was used to calibrate the DSSSD.
A dynamic range of 0--\SI{50}{\mega\electronvolt} is needed for the identification of the implanted {\nitrogen} ions. At the same time, good energy resolution is desirable for the measurement of the $\alpha$ spectrum in the energy range 1.0--\SI{3.5}{\mega\electronvolt}. 
To meet both requirements, the pre-amplifier signal was split and fed to two amplifier-ADC chains with a difference in gain of a factor of 10. The low-gain data is used for the identification of the implanted {\nitrogen} ions, while the high-gain data is used for the $\alpha$ decay spectroscopy. The energy resolution (FWHM) achieved is $\SI{30}{\kilo\electronvolt}$ at \SI{2.4}{\mega\electronvolt} and $\SI{0.9}{\mega\electronvolt}$ at \SI{33}{\mega\electronvolt}.

The primary beam was operated in on/off mode, with the beam gate open for \SI{15}{\second} and then closed for \SI{15}{\second}. A logic signal, representing the state of the beam gate, was fed to the data acquisition. A clean decay spectrum can then be obtained by only including the data collected during the beam-off periods. The data acquisition was triggered by a logic OR between signals in the DSSSD, the $\Delta E$ detector, and the two NaI detectors.

\section{Data reduction}

\subsection{Event reconstruction}\label{sec:sharing}
Energy matching of the signals from the front and back side of the DSSSD allows efficient suppression of electronic noise.
It also allows us to disentangle random coincidences, which occur with significant probability during beam-on periods due to the rather high implantation rate of 10--\SI{20}{\kilo\hertz}.
The condition $\vert E_|front| - E_|back| \vert < \SI{2.5}{\mega\electronvolt}$ is imposed for signals in the low-gain chain and $\vert E_|front| - E_|back| \vert < \SI{0.25}{\mega\electronvolt}$ for signals in the high-gain chain. 
%

%The phenomenon of {\it sharing} complicates energy matching:
%
The DSSSD has a strip width of \SI{300}{\micro\meter} and an interstrip gap of \SI{35}{\micro\meter}, which means that, from purely geometric considerations, \SI{20}{\percent} of the detector surface consists of interstrip regions (\SI{10}{\percent} on each side). In these regions the free charge carriers created by an ionizing particle are unlikely to be collected on a single strip, but are instead shared between the two strips bordering the interstrip region. Therefore, when we detect coincident signals from adjacent strips, we must consider the possibility that the two signals were created by a single particle. If the combined energy matches the energy measured in a strip on the opposite side of the detector, we assume that charge sharing occurred.
%
%Our data show that \SI{18}{\percent} of the $\alpha$ decays and \SI{25}{\percent} of the {\nitrogen} implantations suffer from charge sharing, in good agreement with the geometric estimate of \SI{20}{\percent}. The difference in sharing probability for $\alpha$ decays and {\nitrogen} implantations is not surprising since the free charge carriers created in the two cases have different spatial distributions~\cite{torresi2013}.

Our data show that \SI{8}{\percent} of the $\alpha$ decays suffer from charge sharing between front strips and \SI{10}{\percent} from charge sharing between back strips, in good agreement with the geometric estimate. For the {\nitrogen} implantations we find that \SI{5}{\percent} of the events suffer from charge sharing between front strips while \SI{20}{\percent} suffer from charge sharing between back strips. 
A similar front-back asymmetry was also found by Torresi {\it et al.}\ in a dedicated study of charge sharing in DSSSDs~\cite{torresi2013}, using ions with $Z$-values and energies similar to the {\nitrogen} ions in the present study. Unlike us, however, Torresi {\it et al.}\ find that the probability for sharing between front strips is in agreement with the geometric value. Furthermore, they find that a significant fraction of the sharing events on the front side are associated with opposite-polarity pulses, implying that they cannot be identified by the method used here, which could explain why our value is a factor of 2 short of the geometric estimate. We thus consider it possible that we fail to identify \SI{5}{\percent} of the {\nitrogen} implantations, and we include this as a systematic uncertainty on our final result.%%%OK

\subsection{Identification of implanted \texorpdfstring{{\nitrogen}}{16N}}\label{sec:n16id}
\begin{figure}[htbp]
\centering
\includegraphics[scale=1]{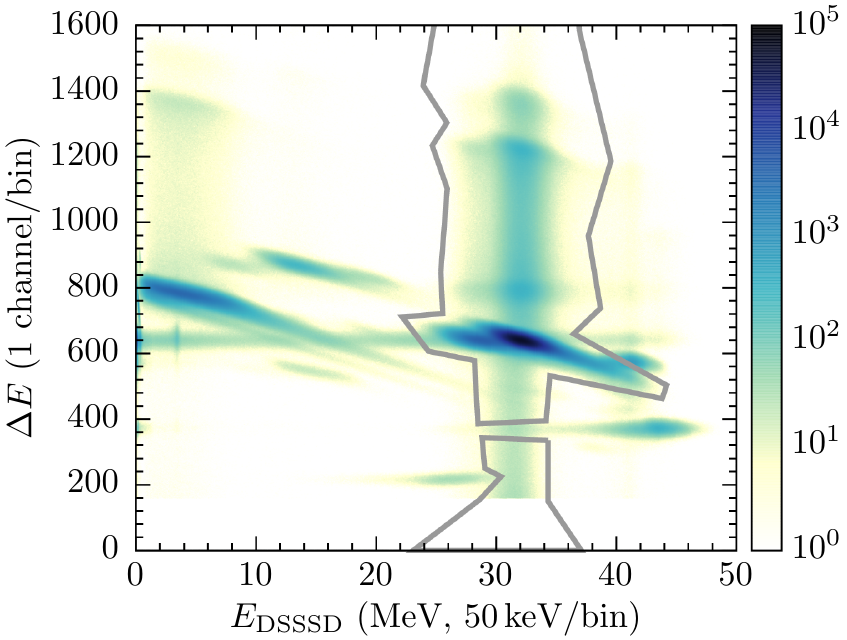}
\caption{Energy signal from the $\Delta E$ detector vs.\ the energy signal from the DSSSD. The grey contour shows the cut used to identify {\nitrogen} implantations.}
\label{EdeltaE:fig}
\end{figure}

%
%The Tri$\mathrm{\mu}$p separator is tuned to select $^{16}\mathrm{N}^{7+}$ ions with an energy of \SI{80}{\mega\electronvolt}. %
%
Using standard methods~\cite{tarasov2008}, the energy loss and straggling of \SI{80}{\mega\electronvolt} $^{16}\mathrm{N}^{7+}$ ions punching through the $\Delta E$ detector are determined to be \SI{47}{\mega\electronvolt} and \SI{3}{\mega\electronvolt}, respectively, leaving on average \SI{33}{\mega\electronvolt} to be deposited in the DSSSD. %
{\nitrogen} implantations are identified by applying appropriate cuts to the $(E_|DSSSD|,\Delta E)$-values, as shown in \fref{EdeltaE:fig}. The vast majority of the {\nitrogen} implantations are contained in the main locus at $E_|DSSSD|\approx\SI{33}{\mega\electronvolt}$ and $\Delta E \approx 700$. %
Pile-up in the $\Delta E$-detector produces the vertical band {\it above} the main locus. %
The vertical band {\it below} the main locus results from {\nitrogen} implantations that induce only a partial signal in the $\Delta E$ detector because they strike the detector close to the periphery of its active area. %
By including these bands we increase the number of identified {\nitrogen} implantations by approximately \SI{5}{\percent}. %
The {\nitrogen} implantations constitute \SI{83}{\percent} of the total number of counts in \fref{EdeltaE:fig}. Other major beam components are $^{13}\mathrm{C}^{5+}$, $^{14}\mathrm{N}^{6+}$, $^{15}\mathrm{N}^{7+}$, $^{13}\mathrm{C}^{6+}$ and protons. No other charge state of $^{16}\mathrm{N}$ is implanted in the DSSSD.
A class of events with anomalous DSSSD response, but with a $\Delta E$-signal consistent with that expected for {\nitrogen} implantations, have also been identified. The origin of these events, which amount to only \SI{2}{\percent} of the number of identified {\nitrogen} implantations, is not fully understood. We include them in the final result as a systematic uncertainty.%%%OK

\subsection{Identification of \texorpdfstring{$\alpha$}{alpha} decays}\label{sec:alphaid}

\begin{figure}[htbp]
\centering
\includegraphics[scale=1]{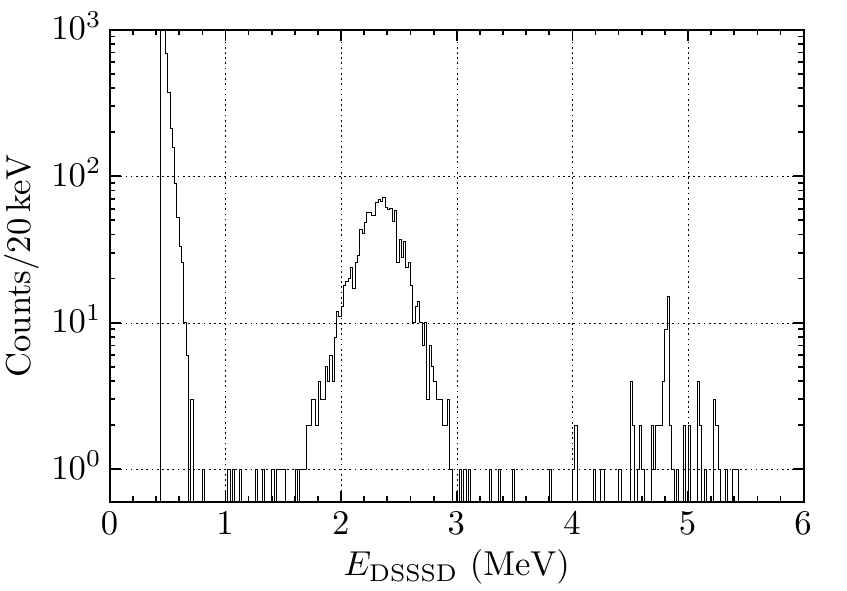}
\caption{Decay spectrum from data taken with the beam gate closed.} 
%Besides the prominent $\alpha$-decay peak we also see some counts at higher energies, probably originating from a contamination of the detector.}
\label{Alphas:fig}
\end{figure}

With a kinetic energy of \SI{33}{\mega\electronvolt} the typical implantation depth of the {\nitrogen} ions in the DSSSD is \SI{27(2)}{\micro\meter} with a straggling of \SI{0.33}{\micro\meter}. Since the $\alpha$ particles from the decay are fully stopped in less than \SI{10}{\micro\meter} of silicon, we can assume full-energy detection for all decays. The decay spectrum obtained during beam-off is practically background-free, see \fref{Alphas:fig}. %2.25MeV alphas have range 8.5um in Si with 0.08um straggling.
The high-energy tail of the $\beta$-particle response extends up to \SI{0.8}{\mega\electronvolt}. Between \SI{1}{\mega\electronvolt} and \SI{3.5}{\mega\electronvolt} we see the signal from the 
$\beta\alpha$ decay of $^{16}$N. The counts above \SI{3.5}{\mega\electronvolt} result from the surface of the DSSSD being contaminated with long-lived $\alpha$-particle emitters (originating from standard calibration sources). From \fref{Alphas:fig} we estimate their contribution below \SI{3.5}{\mega\electronvolt} to be negligible. All counts between 1.0 and \SI{3.5}{\mega\electronvolt} are therefore assumed to be {\nitrogen} $\beta\alpha$ decays. %
We note that the spatial distributions of decay and implantation events are in good agreement, as has been verified by a Kolmogorov-Smirnov test.

\section{Normalization}\label{sec:corr}
The procedure for counting the {\nitrogen} implantations and the $\beta$-delayed $\alpha$ decays has been described in Sections~\ref{sec:n16id} and \ref{sec:alphaid}; however, to determine the branching ratio, $b_{\beta}(9.6)$, we must also correct for the detection efficiency of the experimental setup. Two effects play a main role here: Firstly, we only look for decays when the beam gate is closed, which means that we must apply a correction factor, $C_|on/off|$, that depends on the half-life of {\nitrogen} and the duration of the beam on/off gates. Secondly, there is the issue of the dead time of the data acquisition, which especially affects the data taken with the beam gate open. To determine the necessary correction factors, $C_|dead|^{\mathrm{on}}$ and $C_|dead|^{\mathrm{off}}$, the raw and accepted trigger signals were monitored.

\subsection{Beam-off}%
The total number of decays occurring during the experiment, $N_|decay|^{\mathrm{total}}$, is related to the number of observed decays, $N_|decay|^{\mathrm{obs}}$, by
\begin{align}
N_|decay|^{\mathrm{total}} = C_|dead|^{\mathrm{off}} C_|on/off|^{\phantom{\mathrm{off}}} N_|decay|^{\mathrm{obs}}\, .
\end{align}
Defining the duration of the beam-on gate as $T$ and the duration of the entire beam cycle as $aT$, we can write the analytical expression for the first correction factor as
\begin{align}\label{eq:onoff}
C_|on/off| = \lambda T \frac{1 - e^{-\lambda aT}}{(e^{\lambda T} - 1)(e^{-\lambda T} - e^{-\lambda aT})}\, ,
\end{align}
where $\lambda = \log(2) / t_{\frac{1}{2}}$ is the decay constant. In deriving Eq.~\eqref{eq:onoff} we assume the implantation rate to be constant over a time period of several beam cycles, which is a good approximation in our experiment. Furthermore, we have $t_{\frac{1}{2}}=\SI{7.13 \pm 0.02}{\second}$~\cite{tunl16}, $T = \SI{15}{\second}$ and $a=2$, resulting in the correction factor
\begin{align}
C_|on/off| = \num{2.342 \pm 0.003} \, .
\end{align}
The second correction factor, $C_|dead|^{\mathrm{off}}$, is calculated as the ratio between the number of accepted triggers, $n_|off|$, and the number of raw triggers, $N_|off|$, occurring during the \SI{15}{\second} beam-off period, {\it i.e.}, $C_|dead|^{\mathrm{off}} = N_|off| / n_|off|$. The value of this correction factor varied throughout the experiment between \num{1.011} and \num{1.017}.

\subsection{Beam-on}
The total number of implantations occurring during the experiment, $N_|impl|^{\mathrm{total}}$, is related to the number of observed implantations, $N_|impl|^{\mathrm{obs}}$, by
\begin{align}
N_|impl|^{\mathrm{total}} = C_|dead|^{\mathrm{on}} C_|duty|^{\phantom{on}} N_|impl|^{\mathrm{obs}} \, .
\end{align}
The additional correction factor, $C_|duty|^{\phantom{on}}$, has been introduced to account for a \SI{1}{\kilo\hertz} chopping of the primary beam, necessary for beam intensity monitoring by the cyclotron control system. As a consequence of this chopping, beam is only delivered during a fraction, $D$, of the beam-on period, determined by the duty cycle of the chopper.
Let $N_|on|$ be the number of raw triggers and $n_|on|$ the number of accepted triggers occurring during the \SI{15}{\second} beam-on period, and let $N_|off|$ and $n_|off|$ denote the same quantities for the subsequent \SI{15}{\second} beam-off period. %
We want to determine the number of triggers occuring during the fraction, $D$, of the beam-on period where implantations can take place. 
We assume the trigger rates during the remaining fraction, $(1-D)$, of the beam-on period where the beam is in fact off, to be equal to the trigger rates during the beam-off period. This seems reasonable since the trigger rate during the beam-off period is dominated by the $\gamma$-ray backgound in the NaI scintillators.
Thus, the number of raw and accepted triggers occuring during the fraction, $D$, of the beam-on period where implantations actually take place, may be obtained by subtracting $(1-D)N_|off|$ and $(1-D)n_|off|$ from $N_|on|$ and $n_|on|$, respectively. %
We thus obtain the following expression,
\begin{align}
C_|dead|^{\mathrm{on}} C_|duty|^{\phantom{on}} = \frac{N_|on| - (1-D)N_|off|}{n_|on| - (1-D)n_|off|} \, ,\label{duty:eq}
\end{align}
where $C_|dead|^{\mathrm{on}}=N_|on| / n_|on|$ is the mean dead-time correction.
The convention may seem somewhat artificial, but we keep the mean dead-time correction as a separate factor in order to more easily assess the relative importance of the two effects.
$C_|dead|^{\mathrm{on}}$ is by far the largest correction, varying between \num{1.26} and \num{2.61}, while the additional correction, $C_|duty|$, varies between \num{1.025} and \num{1.087}.
The large variations of $C_|dead|^{\mathrm{on}}$ and, to a lesser degree, $C_|duty|$, are due to significant variations in the beam intensity. The value of $D$ varies between \num{0.22} and \num{0.50}. To give an idea of the relative importance of the various corrections on the final result, we list the values averaged over the entire experiment:
\begin{align}
C_|dead|^{\mathrm{off}} &= \num{1.0136 \pm 0.0014} \nonumber \\
C_|dead|^{\mathrm{on}} &= \num{1.794 \pm 0.007} \nonumber \\
C_|duty| &= \num{1.054 \pm 0.002} \, , \nonumber
\end{align}
where the numbers in parantheses denote the typical statistical uncertainty for a single \SI{30}{\second} beam cycle, which in our view is the most appropriate measure of the overall statistical uncertainty on the correction factors.

\section{Results and discussion}

\begin{figure}[htbp]
\centering
\includegraphics[scale=1]{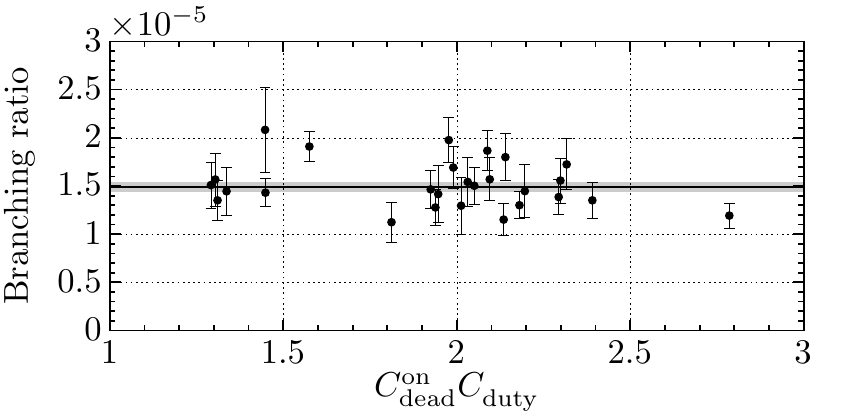}
\includegraphics[scale=1]{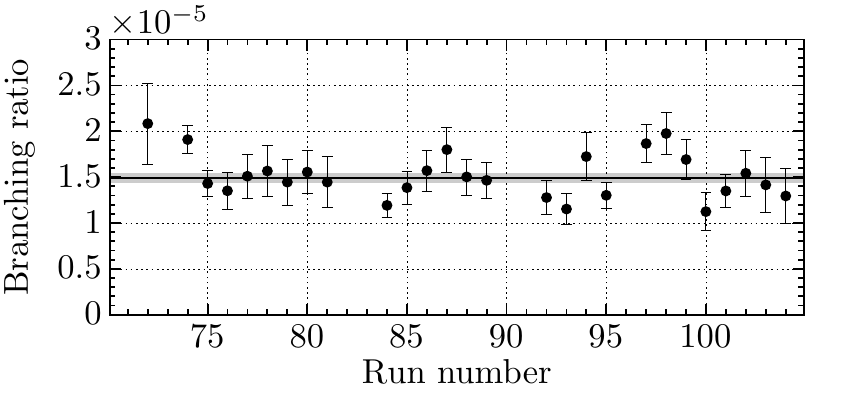}
\caption{\textit{Upper panel:} The $\alpha$-decay branching ratio obtained from the individual runs shown against the value of $C_|dead|^{\mathrm{on}} C_|duty|$, defined in Eq.~\eqref{duty:eq}. \textit{Lower panel:} The $\alpha$-decay branching ratio obtained from the individual runs. The error bars indicate the statistical uncertainty. The solid line shows the weighted mean and the grey area shows the uncertainty on this value.}
\label{result:fig}
\end{figure}

In total 54 hours of data have been collected, split into 27 runs. %
We have identified \num{1.235e8} {\nitrogen} implantations in the beam-on data and 1467 $\alpha$ decays in the beam-off data. The decay spectrum is shown in \fref{Alphas:fig}. The maximum is located at an energy of \SI{2.336 \pm 0.006}{\mega\electronvolt}, where no measures have been taken to correct for $\beta$ summing or the difference in detector response to $\alpha$ particles and $^{12}\mathrm{C}$ ions. %
Taking into account all the correction factors discussed in Section~\ref{sec:corr}, we obtain an $\alpha$-decay branching ratio of $b_{\beta}(9.6)=(1.49\pm 0.04)\times 10^{-5}$. %
Since the correction factor, $C_|dead|^{\mathrm{on}} C_|duty|^{\phantom{on}}$, is quite large, it is also a potential source of significant systematic error. In \fref{result:fig} we show the branching ratio obtained for each of the 27 runs. The error bars indicate statistical uncertainties. The upper panel shows that consistent values are obtained for a rather wide range of correction factors (1.3--2.8). The lower panel shows that the values are constant with time. Thus, we do not find any evidence of systematic errors not accounted for in the dead time analysis. %
The fluctuations around the average are slightly larger than expected from the error bars, resulting in $\chi^2/\mathrm{dof} = 39.3/26$. Because the value of $\chi^2/\mathrm{dof}$ is not quite satisfactory we find it necessary to scale the statistical uncertainty by a factor of $\sqrt{\chi^2/\mathrm{dof}}$. 
Based on the considerations of sections \ref{sec:sharing} and \ref{sec:n16id} the number of {\nitrogen} implantations could be up to \SI{7}{\percent} larger than the number we find, and we include this as a systematic uncertainty, giving a %%%JR
final result for the branching ratio of
\begin{align}
b_{\beta}(9.6) = \left( 1.49\pm 0.05\textrm{(stat)}^{+0.0}_{-0.10}\textrm{(sys)} \right) \times 10^{-5}.
\end{align}
This result represents a \SI{24}{\percent} increase compared to the value of \num{1.20 \pm 0.05 e-5} reported by the Mainz group~\cite{kaufmann1961} and quoted in the most recent TUNL evaluation~\cite{tunl16}. On the other hand, it is consistent with the less precise value of \num{1.3 \pm 0.3 e-5} recently obtained by Zhao {\it et al.}~\cite{zhao1993} using an experimental technique similar to the one described here. %

We have performed an $R$-matrix analysis to estimate the impact on the extrapolated $S$-factor~\cite{azuma2010}. Our analysis suggests an increase of \SI{24}{\percent} in $S_{\textrm{E}1}(0.3)$ and hence an increase of \SI{13}{\percent} in the total $S$-factor if we adopt the same values as in Section~\ref{sec:s-factor}. A similar strong correlation between $b_{\beta}(9.6)$ and $S_|E1|(0.3)$ has also been found by other authors~\cite{azuma1994}. %
It is interesting to note that our new determination of $b_{\beta}(9.6)$ results in a slightly improved agreement between the preferred $S_|E1|(0.3)$ value implied by the $\beta\alpha$-decay data (previously \SI{84}{\kilo\electronvolt\barn}, now \SI{104}{\kilo\electronvolt\barn}) and the $S_|E1|(0.3)$ value implied by the sub-Coulomb $\alpha$-transfer data (\SI{101}{\kilo\electronvolt\barn}~\cite{brune1999}). However, when the error bars are taken into account (\SI{\pm 21}{\kilo\electronvolt\barn} and \SI{\pm 17}{\kilo\electronvolt\barn}, respectively) the improvement cannot be said to be significant.

\section{Conclusion}

We have measured the $\beta\alpha$ branching ratio of $^{16}$N using a technique that avoids the difficulties associated with the absolute counting of $\beta$ particles. We implant a high-energy $^{16}$N beam in a finely segmented Si detector and measure the energy signals from the implantation and the $\alpha$ decay. The branching ratio is then obtained in a straightforward manner as the ratio of $\alpha$ decays to implantations. Our result, $( 1.49\pm 0.05\textrm{(stat)}^{+0.0}_{-0.10}\textrm{(sys)} ) \times 10^{-5}$, represents a \SI{24}{\percent} increase compared to the accepted value~\cite{tunl16}. %
Our branching-ratio determination leads to a $S_{\textrm{E}1}(0.3)$ value that is \SI{24}{\percent} higher than previously believed. Adopting the same values as in Section~\ref{sec:s-factor}, the total $S$-factor goes up by \SI{13}{\percent}. Given the significant astrophysical implications of such a change in the $S$-factor, it would be desirable to have our result confirmed in an independent experiment. Independent confirmation of the $\beta$ branching ratio recently reported by Tang~{\it et al.}\ for the \SI{7.12}{\mega\electronvolt} state~\cite{tang2010}, would also be of interest.

\section*{Acknowledgements}
We thank James de Boer and Karsten Riisager for helpful discussion on the $R$-matrix analysis. OSK acknowledges support from the Villum Foundation. 
We also acknowledge financial support from the European Research Council under ERC starting grant LOBENA, No.\ 307447, the Stichting voor Fundamenteel Onderzoek der Materie (FOM) under Programme 114 (TRI$\mu$P), the European Commission within the Seventh Framework Programme through IA-ENSAR (contract no.\ RII3-CT-2010-262010), FWO-Vlaanderen (Belgium), GOA/2010/010 (BOF KU Leuven) and the Interuniversity Attraction Poles Programme initiated by the Belgian Science Policy Office (BriX network P7/12).

\section*{References}
%%%\bibliography{mybibliography}
\bibliography{okirsebom}
\end{document}